\documentclass[11pt,twoside]{article}

\usepackage{asp2006}
\usepackage{epsf}
\usepackage{lscape}

\begin{document}
\title{Fluorine in R Coronae Borealis and Extreme Helium Stars}   
\author{Gajendra Pandey\altaffilmark{1}, David L.\ Lambert\altaffilmark{2}, and 
N. Kameswara Rao\altaffilmark{1}}   

\altaffiltext{1}{Indian Institute of Astrophysics;
Bangalore,  560034 India}    

\altaffiltext{2}{The W.J. McDonald Observatory, University of Texas at Austin; Austin,
TX 78712-1083}

\begin{abstract} 

Neutral fluorine (F\,{\sc i}) lines are identified in the optical
spectra of several R\,Coronae Borealis stars (RCBs) at maximum light.
These lines provide the first measurement of the fluorine
abundance in these stars. Fluorine is enriched in some RCBs by factors of 800 to 8000
relative to its likely initial abundance.
The overabundances of fluorine are evidence for the synthesis
of fluorine. These results are discussed in the light of the scenario that
RCBs are formed by accretion of an He white dwarf by a C-O white dwarf.
Sakurai's object (V4334\,Sgr), a final He-shell flash product, shows no
detectable F\,{\sc i} lines.

\end{abstract}


\section{Introduction}   

R Coronae Borealis (RCB) stars comprise a sequence of hydrogen-deficient
supergiants with effective temperatures from about $T_{\rm eff} = 3500$ K,
as represented by Z\,UMi and DY\,Per, to about 19,500 K, as represented
by DY\,Cen.
The characteristic of H-deficiency is shared by the H-deficient cool
carbon (HdC) stars at low temperatures and by the extreme helium
(EHe) stars at high temperatures. A common assumption is that the
sequence HdC - RCB - EHe in the ($T_{\rm eff},\log g$) plane
reflects a close evolutionary connection. In this sequence, the
RCBs are distinguished from HdC and EHe stars by their second principal defining
characteristic: their propensity to fade, in visual light, unpredictably as a
cloud of carbon dust obscures the star. This propensity is not universally shared:
XX\,Cam has yet to be observed below maximum light. In addition, DY\,Cen might be
considered as an EHe star known to experience RCB-like fadings.

If HdC, RCB, and EHe stars share a common heritage, the expectation
is that their atmospheric compositions should show some common
features \citep{pan2004a,rao2005}. It is through the compositions that one hopes to test
theoretical ideas about the origins of these extremely rare
stars; just five HdC, about 40 RCB \citep{zanie2005}, and 21 EHe
stars are known in the Galaxy. Currently, two
scenarios remain in contention to account for these
H-poor high luminosity stars. In the first, a final He-shell flash in a
post-AGB star on the white dwarf cooling track creates a H-poor luminous
star. This is dubbed the `final flash' (FF) scenario. In the second, the H-poor
star is formed from a merger of a He white dwarf with a C-O white dwarf.
In a close binary system, accretion of the He white dwarf by the C-O white dwarf
may lead to a H-poor supergiant with the C-O white dwarf as its core. This is
called the `double degenerate' (DD) scenario.

Products of both the FF and DD scenarios may be presumed to exist.
A determination of which scenario provided which star rests in large
part on the observed chemical composition of a star's atmosphere and theoretical
predictions about the FF and DD products. Evidence from
elemental abundances, especially the H, C, N, and O abundances, suggests that
the RCB and EHe stars evolved via the
DD rather than the FF route \citep{asp00,pan2001,saio02,pan2006}.
Convincing, essentially incontrovertible, evidence
that the DD scenario led to the HdCs and some cool RCBs was
presented by \citet{clay2007} with their discovery that
the $^{18}$O was very abundant in their atmospheres. This usually
rare isotope of oxygen was attributed to nucleosynthesis occurring
during and following accretion of the He-rich material onto the C-O white dwarf.

Determination of the oxygen isotopic ratios demands
a cool star with the CO vibration-rotation bands in its spectrum.
The majority of RCBs and all of the EHes are too hot for CO to contribute
to their spectra \citep{tenen2005}.
An alternative tracer of nucleosynthesis during a merger may
be provided by the fluorine abundance. Considerable enrichment
of EHe stars with F was discovered by \citet{pana2006} from detection and
analysis of about a dozen F\,{\sc i} lines in optical spectra of
cool EHe stars. Clayton et al.'s (2007) calculations suggest that F synthesis is
possible in the DD scenario. Here, we report on a search for
F\,{\sc i} lines in spectra of RCBs and discuss the F abundances in
light of the results for EHes and the expectations for the DD and FF
scenarios.

\section{Observations}

High-resolution optical spectra of RCBs at maximum light obtained
at the W.J. McDonald Observatory and the Vainu Bappu Observatory
were examined for the presence of  F\,{\sc i} lines.

Suitable McDonald spectra of 13 RCBs were available for analysis. In
several cases, spectra of the same star from different epochs were
available. The stars are listed in Table 1.
At VBO, spectra were obtained of UW\,Cen, a star inaccessible from
the McDonald Observatory, and also of XX\,Cam, R\,CrB, RY\,Sgr, and VZ\,Sgr.
The UW\,Cen spectrum was complemented by  a CTIO spectrum from 1992 (see
Asplund et al. 2000).

\section{F\,{\sc i} Lines}

In the atmosphere of a RCB star, fluorine is present as neutral
atoms. The leading lines of F\,{\sc i} in optical spectra come from excited levels
with lower excitation potentials of 12.7 eV or higher. 
The adopted $gf$-values come from \citet{musie99}.
The following F\,{\sc i} lines were  identified as the principal or
leading contributor to a stellar line: 7398.68, 7754.69, 6902.47, 7425.65,
and 6834.26\AA.
The measured mean fluorine abundances (Table 1) are given as
log $\epsilon(\rm F)$, normalized such that log $\Sigma$$\mu_i \epsilon(i)$ = 12.15
where $\mu_i$ is the atomic weight of element $i$.
The errors of the derived  F abundances
given in Table 1 are the line-to-line scatter. 

\begin{landscape}
\begin{table}[!ht]
\caption{The analyzed RCBs, their stellar parameters,
and fluorine abundances from individual F\,{\sc i} lines. The Sakurai's object, a final He-shell flash product, is also listed.}
\smallskip
\begin{center}
{\small
\begin{tabular}{llcccccr}
\tableline
\noalign{\smallskip}
Star & ($T_{\rm eff}$, $\log g$, $\xi$) & \multicolumn{6}{c}{log $\epsilon(\rm F)$}\\
\noalign{\smallskip}
\cline{3-8} 
\noalign{\smallskip}
  &   & 7398.68\AA\ & 7754.69\AA\ & 6902.47\AA\ & 7425.64\AA\ & 6834.26\AA\ & Mean \\
\noalign{\smallskip}
\tableline
\noalign{\smallskip}
V3795\,Sgr & (8000, 1.0, 10.0) & ... & 6.60 & 6.65 & 6.70 & 6.70 & 6.66$\pm$0.05(4)\\
&&&&&&&\\
UW\,Cen & (7500, 1.0, 12.0) & 7.20 & ... & 7.00 & 7.10 & 7.20 & 7.1$\pm$0.1(4)\\
&&&&&&&\\
RY\,Sgr & (7250, 0.75, 6.0) & ... & ... & 6.80 & ... & 7.10 & 6.95$\pm$0.2(2) \\
&&&&&&&\\
XX\,Cam & (7250, 0.75, 9.0) & ... & ... & $<$5.6 & ... & $<$5.6 & $<$5.6 \\
&&&&&&&\\
UV\,Cas & (7250, 0.5, 7.0) & ... & ... & 6.20 & ... & ... & 6.2(1)\\
&&&&&&&\\
UX\,Ant & (7000, 0.5, 5.0) & ... & ... & $<$6.2 & ... & $<$6.2 & $<$6.2\\
&&&&&&&\\
VZ\,Sgr & (7000, 0.5, 8.0) & ... & ... & 6.30 & ... & 6.50 & 6.4$\pm$0.1(2)\\
&&&&&&&\\
R\,CrB & (6750, 0.5, 7.0) & ... & ... & 6.85 & ... & 7.00 & 6.9$\pm$0.1(2)\\
&&&&&&&\\
V2552\,Oph & (6750, 0.5, 7.0) & ... & ... & 6.60 & ... & ... & 6.6(1)\\
&&&&&&&\\
V854\,Cen & (6750, 0.0, 6.0) & ... & ... & ... & ... & $<$5.7 & $<$5.7\\
&&&&&&&\\
SU\,Tau & (6500, 0.5, 7.0) & ... & ... & 6.90 & ... & 7.00 & 6.95$\pm$0.1(2)\\
&&&&&&&\\
V\,CrA & (6500, 0.5, 7.0) & ... & ... & 6.5: & ... & ... & 6.5:(1)\\
&&&&&&&\\
V482\,Cyg & (6500, 0.5, 4.0) & ... & ... & ... & ... & 6.6: & 6.6:(1)\\
&&&&&&&\\
GU\,Sgr & (6250, 0.5, 7.0) & ... & ... & ... & ... & 7.2: & 7.2:(1)\\
&&&&&&&\\
FH\,Sct & (6250, 0.25, 6.0) & ... & ... & ... & ... & 7.2: & 7.2:(1)\\
&&&&&&&\\
Sakurai's object & (7500, 0.0, 8.0) & ... & ... & ... & ... & $<$5.4 & $<$5.4\\
&&&&&&&\\
\noalign{\smallskip}
\tableline
\end{tabular}
}
\end{center}
\end{table}
\end{landscape}

\section{Discussion}

The analyzed RCBs (excluding stars with upper limits to the F abundances and,
uncertain F abundances) have a mean F abundance of 6.7 which is the same
for the analyzed EHes \citep{pana2006}. Thus, the F abundances are similar across
an effective temperature range from about 6500 K to 14000 K, an indication
that non-LTE effects are possibly small.
 The F abundances of the analyzed RCB and
EHe stars show no obvious trend with their abundances of other elements.
More interestingly, the F overabundances are extremely large:
enhancements of about 800, 2500, and 8000 at $\log\epsilon$(Fe)$ = 7.5$, 6.5, and 5.5,
respectively.

\subsection{The DD Scenario and Fluorine}

In the `cold' (i.e., no nucleosynthesis during the merger) version of the
DD scenario, the He white dwarf is stripped and accreted by the
C-O white dwarf. In the `hot' DD scenario, nucleosynthesis
occurs during and following accretion. 
Fluorine in PG1159 stars shows a range of abundances \citep{werner2005,werner2006}
from solar to 250 times solar. As Werner \& Herwig discuss, this
range is not out of line with theoretical predictions for the He-intershell.
But in the cold DD scenario, the He-intershell material
is diluted, according to the canonical recipe, by a factor of about ten and,
then, overabundances of up to 25 times solar for the
RCBs and EHes are predicted. The observed overabundances of F range upward of 1000 times.
$^{19}$F synthesis is demonstrated by \citet{clay2007} which was briefly about 100 times
above its solar abundance. Challenge is to show that the `hot' DD scenario
includes the possibility of robustly increasing the F abundances to the observed
levels of 1000 times over solar. 

\acknowledgements 
GP thanks Klaus Werner, Thomas Rauch, the Organising committee, and IIA Bangalore
for all their support.


\end{document}